\documentclass[11pt,twoside,a4paper,cmspaper,final,collab]{cms-tdr}
\def\svnVersion{ecfb7b2}\def\svnDate{2026/07/22}\def\cmsCernNoTag{CERN-EP-2026-004}\def\cmsCernDate{\today}\def\cmsMessage{Published in the Journal of High Energy Physics as \href{http://dx.doi.org/10.1007/JHEP07(2026)106}{\doi{10.1007/JHEP07(2026)106}.}}
\begin{document}\cmsNoteHeader{BPH-24-004}

\newcommand{\QQbar}{\ensuremath{\PQQ\PAQQ}\xspace}
\newcommand{\jpsi}{\JPsi}
\newcommand{\psip}{\ensuremath{\psi\mathrm{(2S)}}\xspace}
\newcommand{\tinynegspace}{{\kern-1pt}}
\newcommand{\oneSzero}{\ensuremath{{}^1{\tinynegspace}S_0^{[8]}}\xspace}
\newcommand{\threeSone}{\ensuremath{{}^3{\tinynegspace}S_1^{[8]}}\xspace}
\newcommand{\threePJ}{\ensuremath{{}^3{\tinynegspace}P_J^{[8]}}\xspace}
\newcommand{\UpsilonOneS}{\ensuremath{\PgUa}\xspace}
\newcommand{\UpsilonTwoS}{\ensuremath{\PgUb}\xspace}
\newcommand{\UpsilonThreeS}{\ensuremath{\PgUc}\xspace}
\newcommand{\UpsilonnS}{\ensuremath{\PgU(n\mathrm{S})}\xspace}

\cmsNoteHeader{BPH-24-004}
\title{Measurement of the \UpsilonOneS, \UpsilonTwoS, and \UpsilonThreeS differential cross sections in $\Pp\Pp$ collisions at \texorpdfstring{$\sqrt{s} = 13.6\TeV$}{sqrt(s) = 13.6 TeV}}

\author[cern]{The CMS Collaboration}

\date{\today}

\abstract{The production cross sections of the \UpsilonOneS, \UpsilonTwoS, and \UpsilonThreeS mesons are measured in proton-proton collisions at $\sqrt{s}=13.6$\TeV, using a data sample collected in 2022 by the CMS experiment and corresponding to an integrated luminosity of 37.4\fbinv. 
The measurement is performed in the $\PGmp\PGmm$ decay channels, differentially as a function of transverse momentum in the 20--200\GeV range, in the $\abs{y} < 0.6$ and $0.6 < \abs{y} < 1.2$ rapidity intervals.}

\hypersetup{%
pdfauthor={CMS Collaboration},%
pdftitle={Measurement of the Upsilon(1S), Upsilon(2S), and Upsilon(3S) differential production cross sections in pp collisions at sqrt(s) = 13.6 TeV},%
pdfsubject={CMS},%
pdfkeywords={CMS, cross sections, quarkonium production, Upsilon}
}

\maketitle 

\section{Introduction}

The production of charmonium and bottomonium states 
(bound states of, respectively, charm and beauty quark-antiquark pairs)
provides crucial insights for the understanding of hadron formation 
in the scope of quantum chromodynamics (QCD), the theory of strong interactions~\cite{QWG:2004,Brambilla}. 
Several theoretical models have been developed to describe quarkonium production, 
including the ``colour singlet model" and the ``colour evaporation model" 
(Refs.~\cite{CEM,Lansberg} and references therein).
Currently, the most widely accepted theoretical description of quarkonium production is based on the assumption 
that the charm and beauty quarks are sufficiently heavy to allow the factorization of short- and long-distance effects,
a conjecture at the basis of the non-relativistic QCD (NRQCD) framework~\cite{NRQCD}.
Within this theoretical approach,
quarkonium production is described as a superposition of several contributions,
each corresponding to a possible configuration of the intermediate $\QQbar(^{2S+1}L_J^{[c]})$ state, 
characterized by its spin ($S$), orbital angular momentum ($L$), total angular momentum ($J$), 
and colour ($c = 1$ or 8).
These contributions are 
organized in an expansion in powers of the relative \QQbar velocity
so that only a small number of leading and subleading terms remain quantitatively relevant.
In the case of S-wave quarkonia, 
the leading contributions are the \oneSzero, \threeSone, and \threePJ colour octets~\cite{QWG:2004,Brambilla,NRQCD},
the first of them being unpolarized.
Each term contributes to the quarkonium production cross section as the product of two factors.
The first factor is the production cross section of the \QQbar ``pre-resonance", 
a function of its transverse momentum (\pt) and rapidity ($y$) evaluated through perturbative QCD computations
and denoted as the ``short-distance coefficient" (SDC).
The second, known as the ``long-distance matrix element" (LDME),
is a phenomenological parameter (supposedly constant and universal) 
reflecting the probability of the (non-perturbative) evolution of that specific \QQbar state to the observed meson.
For S-wave charmonium and bottomonium production, 
the dominant octet terms have been computed up to next-to-leading order (NLO) 
accuracy in the strong coupling parameter~\cite{Butenschoen:2012qh,Chao:2012iv,Butenschoen:2012qr,Gong:2012ug}.
The LDMEs are not calculable by the currently existing theory methods and are, 
instead, determined from global fits to experimental data.
Different sets of LDMEs have been extracted in the past, 
depending on which sets of measurements are included in the fit,
on the minimum \pt threshold used in the data selection, and, most significantly, 
on whether the fit analyses are restricted to differential cross sections 
or also include polarization measurements.
More details can be found in Refs.~\cite{Faccioli:EPJC69,Butenschoen:2012qh,Butenschoen:2012qr,
Chao:2012iv,Gong:2012ug,Butenschoen:2012px,Faccioli:PLB736,
Bodwin:2015iua,Faccioli:EPJC78,Wang:2012gf,Gong:2013nloUpsilon,Han:2014UpsilonChib}
and references therein.

Production cross sections of S-wave quarkonia have been measured
in proton-proton ($\Pp\Pp$) collisions at the LHC,
as functions of \pt and in several $y$ ranges, by the ALICE, ATLAS, CMS, and LHCb
Collaborations~\cite{CMS:2010nis,BPH-10-004,LHCb:2011zfl,ALICE:2011zqe,ATLAS:2014zpz,BPH-14-001,
LHCb:2015foc,ATLAS:2015zdw,ATLAS:2023qnh,
CMS:2010wld,LHCb:2012aa,ATLAS:2012lmu,CMS:2013qur,CMS:2015xqv,CMS13TeV}.
Quarkonium polarization measurements have also been performed,
providing complementary information regarding the production 
mechanisms~\cite{ALICE:2011gej,BPH-13-003,CMS:BPH-11-023,LHCb:2013izl,LHCb:2014brf,CMS:2019jas,BPH-22-009}.
More details on these measurements can be found in Ref.~\cite{sym17091521} and references therein.
This article reports a measurement of the \UpsilonOneS, \UpsilonTwoS, and \UpsilonThreeS
(denoted by \UpsilonnS) differential cross sections
in $\Pp\Pp$ collisions at $\sqrt{s} = 13.6$\TeV,
on the basis of a data sample collected in 2022 by the CMS experiment,
corresponding to an integrated luminosity of 37.4\fbinv.
The measurements are performed in the dimuon decay channels
and cover the broad 20--200\GeV \pt range,
in the $\abs{y} < 0.6$ and $0.6 < \abs{y} < 1.2$ regions.
The reported cross sections include feed-down contributions from heavier bottomonia.

With respect to the results obtained with a $\sqrt{s} = 13$\TeV $\Pp\Pp$ data sample
reported in a previous CMS publication~\cite{CMS13TeV},
this analysis extends the \pt reach from 100 to 200\GeV. 
The \UpsilonnS signal yields, $N(\pt,\abs{y})$, are extracted, 
in each $(\pt,\abs{y})$ two-dimensional bin,
from an extended binned maximum likelihood fit to the dimuon invariant mass spectrum. 
The two-dimensional acceptance, $\mathcal{A}(\pt, \abs{y})$,
is computed through detector simulation, 
while the corresponding detection efficiencies, $\epsilon(\pt, \abs{y})$, are evaluated 
from measured data using the tag-and-probe method~\cite{CMS:2010wld, CMS13TeV, CMS:2010nis} for the single muon terms
and from the simulated event samples for the terms that depend on the muon-pair correlations. 
All measurements are reported in terms of the product between the 
\UpsilonnS production cross section, $\sigma$,
and the branching fraction of the corresponding dimuon decay channel, 
$\mathcal{B}(\PgU \to \PGmp\PGmm)$, abbreviated as $\mathcal{B}$ in the following.
In summary, 
we compute the \pt-differential cross section times branching fraction, 
per unit of rapidity and in two rapidity ranges, as
\begin{equation}
\label{crosssection}
\mathcal{B} \, \frac{\rd^2 \sigma}{\rd y \rd\pt} =
\frac{N(\pt,\abs{y})}{\mathcal{L} \, \Delta y \, \Delta \pt} \,
\frac{1}{\epsilon(\pt, \abs{y})} \,
\frac{1}{\mathcal{A}(\pt, \abs{y})} \, ,
\end{equation}
where $\mathcal{L}$ is the integrated luminosity, 
$\Delta \pt$ is the width of the \pt bin under consideration,
and $\Delta y = 1.2$ for both $\abs{y}$ bins.
The numerical values are reported in the HEPData record of this analysis~\cite{HEPData}.

\section{The CMS apparatus, event samples, and selection criteria}

The CMS apparatus is a multipurpose detector~\cite{CMS:2008xjf}
designed to trigger on and identify electrons, muons, photons,
and (charged and neutral) hadrons~\cite{CMS:2020uim,CMS:2018rym,CMS:2014pgm}.
A superconducting solenoid of 6\unit{m} internal diameter provides a magnetic field of 3.8\unit{T}.
Within the solenoid volume are the silicon pixel and strip tracker,
a crystal electromagnetic calorimeter, and a brass and scintillator hadron calorimeter.
Forward calorimeters extend the pseudorapidity ($\eta$) coverage provided by the barrel and endcap detectors.
Muons are measured in the $\abs{\eta} < 2.4$ range,
with gas-ionization detectors 
(drift tubes, cathode strip chambers, and resistive plate chambers)
embedded in the steel flux-return yoke outside the solenoid.
Muons are identified as tracks in the silicon tracker consistent with 
either a track or several hits in the muon system, 
and associated with calorimeter deposits compatible with the muon hypothesis. 
The relative \pt resolution is 1\% in the barrel and 3\% in the endcaps, for muons with \pt up to 100\GeV. 
More detailed descriptions of the CMS detector, 
together with a definition of the coordinate system used and the relevant kinematic variables, 
can be found in Refs.~\cite{CMS:2008xjf,CMS:2023gfb}.

The events of interest are selected using a two-tiered trigger system.
The first level (L1), composed of custom hardware processors,
uses information from the calorimeters and muon detectors to select events 
at a rate of 100\unit{kHz} within a fixed latency of 4\mus~\cite{CMS:2020cmk}.
The L1 trigger used in this analysis 
selects events with at least two opposite-sign muons with $\pt > 4.5$\GeV and $\abs{\eta}<2$,
having a dimuon mass, $m_{\PGm\PGm}$, larger than 7\GeV. 
The second level,
consisting of a farm of processors running a faster version of the full event reconstruction software,
reduces the rate to around 1\unit{kHz} before data storage~\cite{CMS:2016ngn,CMS:2024aqx}.
The trigger used in this analysis 
only retains dimuons with $8.5 < m_{\PGm\PGm} < 11.5\GeV$,
$\pt > 9.9$\GeV, and $\abs{y} < 1.4$. 
The two muon tracks are fitted together, imposing a common vertex, 
and the $\chi^2$ fit probability is required to be larger than 0.5\%.
The events selected for the final analysis also need to pass data-quality certification criteria.

At the offline analysis level, 
where the kinematical properties of the muons are known with a higher precision than at the trigger level, 
further selection criteria are applied.
To start with, the selected muons must match the trigger requirements and 
pass certain (``soft muon") quality requirements~\cite{CMS:2018rym}.
Furthermore, only muons that have $\pt > 10$\GeV and $\abs{\eta} < 1.4$ are kept.
In addition, the dimuon vertex fit is redone and the resulting $\chi^2$ probability is required to be larger than 1\%.
If an event has more than one muon pair passing the requirements 
(happening in less than 0.05\% of the selected events), 
only the pair with the largest vertex $\chi^2$ probability is kept.

For several steps of the analysis procedure, described below,
detailed Monte Carlo (MC) simulations are performed,  
covering the full data collection and reconstruction chain, 
from the trigger step to the offline event selection.
The event samples are generated 
with the \PYTHIA\,8.306 event generator~\cite{PYTHIA,Pythia83},
configured with the CP5 tune~\cite{CMS:CP5tune} 
and the NNPDF3.1 NNLO set of parton distribution functions~\cite{NNPDF:2017mvq}. 
The emitted muons undergo final-state radiation, generated through the \PHOTOS\,3.61 package~\cite{PHOTOS2}.
The simulation includes effects from multiple $\Pp\Pp$ interactions in the same or nearby bunch crossings, 
with a multiplicity distribution matching the data.
The events are then processed through a 
simulation of the CMS detector,
based on the \GEANTfour package~\cite{GEANT4}, 
using the same trigger, reconstruction algorithms, and selection criteria as used to collect and process the data.

\section{Acceptance and efficiencies}

The acceptances, $\mathcal{A}(\pt, \abs{y})$, 
are calculated using a large sample of simulated \UpsilonOneS events, 
covering a broader phase space window than that used in the analysis.
The \UpsilonOneS \pt and $\abs{y}$ distributions used in the event generation 
are reweighted by the ratios between the background-subtracted measured spectra~\cite{sPlot} 
and the simulated spectra.
For each $(\pt,\abs{y})$ bin of the \UpsilonOneS, 
the acceptance is computed as the fraction of generated events 
where both decay muons have $\pt > 10$\GeV and $\abs{\eta} < 1.4$.
The differences between the acceptance values evaluated for the three bottomonium states 
are negligible (less than 1\%), given the high granularity of the acceptance tables, 
as well as their similar masses and decay kinematics, 
so that the \UpsilonOneS values are also used to obtain the 
\UpsilonTwoS and \UpsilonThreeS cross sections.

The default acceptances are computed assuming that the \UpsilonnS mesons are produced unpolarized, 
an assumption compatible with the currently available measurements~\cite{CMS:BPH-11-023}.
The reported cross sections can be converted to those that would have been obtained
if other polarization scenarios had been assumed, by using the correction factors 
presented in Table~\ref{tab:polarization-factors} of Appendix~\ref{app:tables},
corresponding to full transverse or full longitudinal production
(\ie, assuming that the polar asymmetry parameter in the centre-of-mass helicity frame
takes the extreme values $\lambda_\vartheta = \pm 1$).

The detection efficiencies of the muons used in the analysis are factorized into three terms:
trigger, muon identification, and muon reconstruction.
They have been measured, as functions of the muon \pt and in several $\abs{\eta}$ bins,
with a tag-and-probe technique~\cite{CMS:2010wld, CMS:2010nis, CMS13TeV}
applied to a sample of \UpsilonOneS events collected with single muon triggers.
The dimuon efficiency that enters in Eq.~(\ref{crosssection}) is computed 
as the product of the two single-muon efficiencies, 
times a scaling factor that corrects for the detection inefficiency 
that occurs when the two muons are emitted close to each other in angular and momentum space,
so that they might be considered as a single track at L1 and, hence, 
fail to trigger the data acquisition system~\cite{BPH-10-004, BPH-13-003, BPH-14-001, BPH-22-009}.
This correction factor, which also accounts for the dimuon vertex inefficiency,
is evaluated using the detailed simulation of the full detection chain.

\section{Yield extraction}

The \UpsilonnS signal yields are evaluated, in each $(\pt,\abs{y})$ bin,
by performing a binned extended maximum likelihood fit (using RooFit~\cite{Verkerke:2003ir}) 
to the measured dimuon invariant mass distribution.
The measured distributions are fitted with a model composed of three signal terms, 
each of them parametrized by the sum of two Crystal Ball (CB) functions~\cite{CB},
plus a second-order polynomial function to describe the underlying continuum background.
The two CB functions that describe each of the three $\PgU$ resonances
have shapes determined by their means ($\mu_1$ and $\mu_2$),
widths ($\sigma_1$ and $\sigma_2$), 
and non-Gaussian tail parameters ($n_1$, $n_2$, $\alpha_1$, and $\alpha_2$).
This fit model (and the fixed shape parameters mentioned below) 
has been defined from studies of simulated event distributions, 
validated with data distributions integrated over \pt.

\begin{figure}[ht]
\centering
\includegraphics[width=0.46\linewidth]{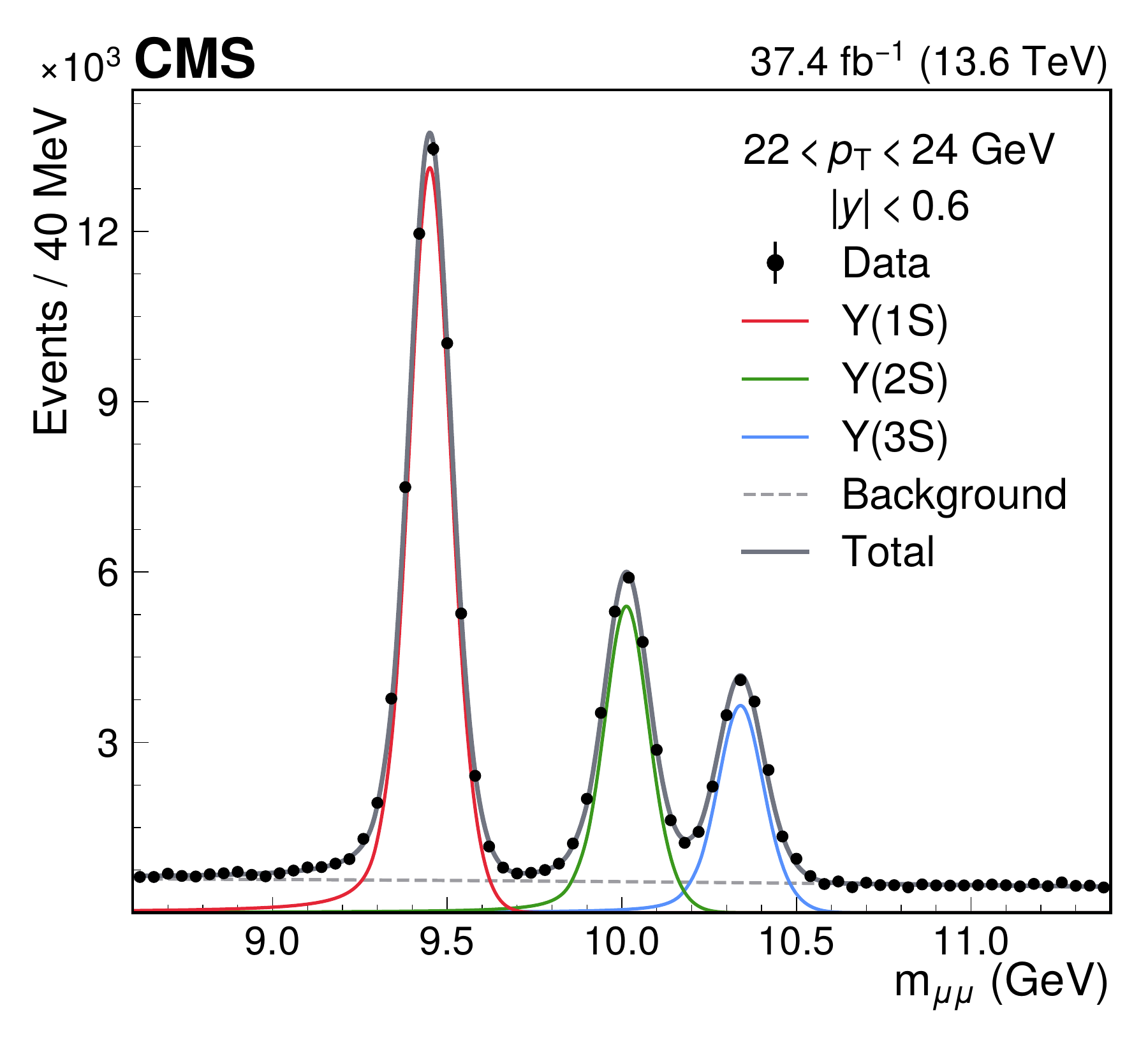}
\includegraphics[width=0.46\linewidth]{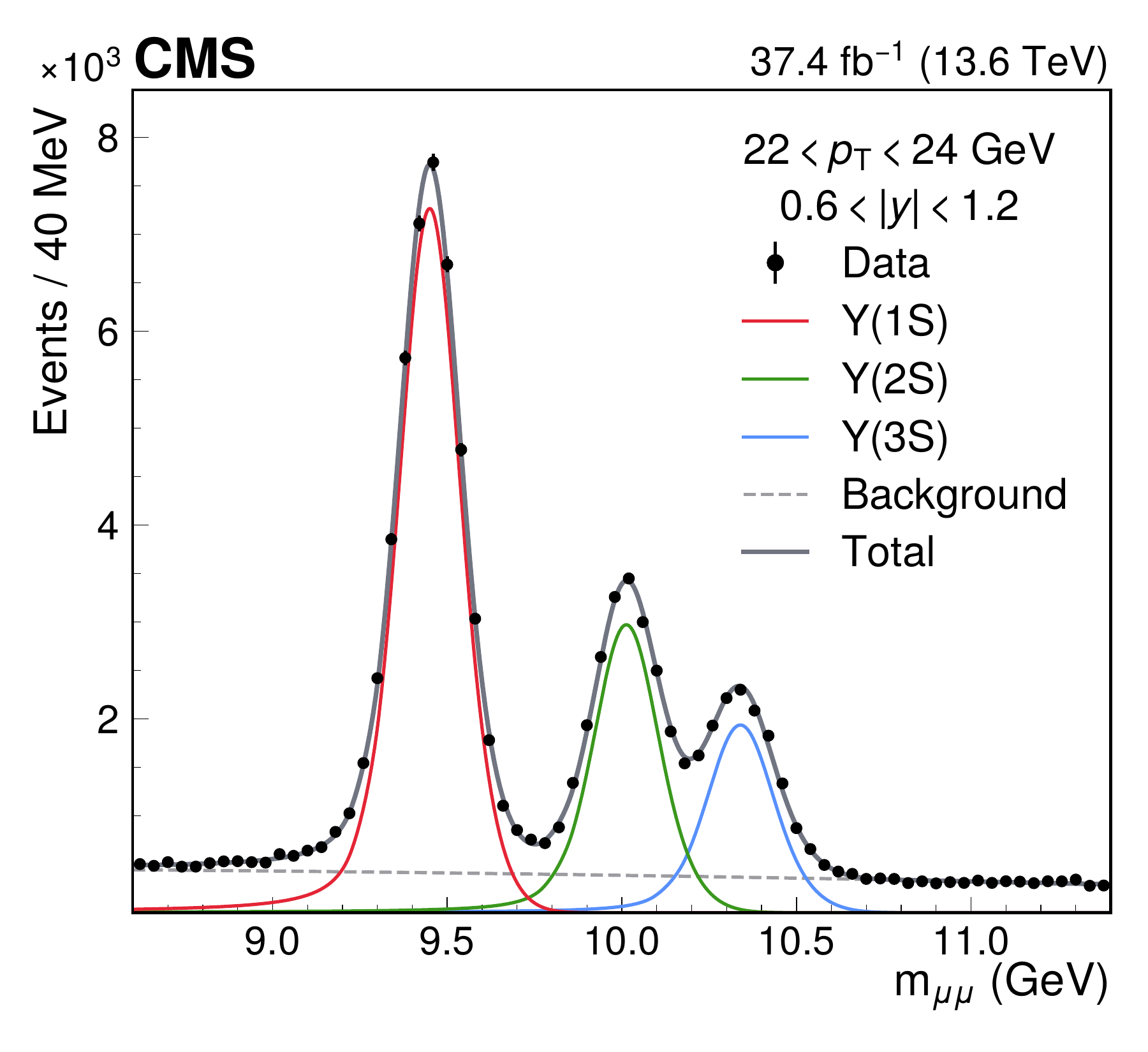}
\vglue-2mm
\includegraphics[width=0.46\linewidth]{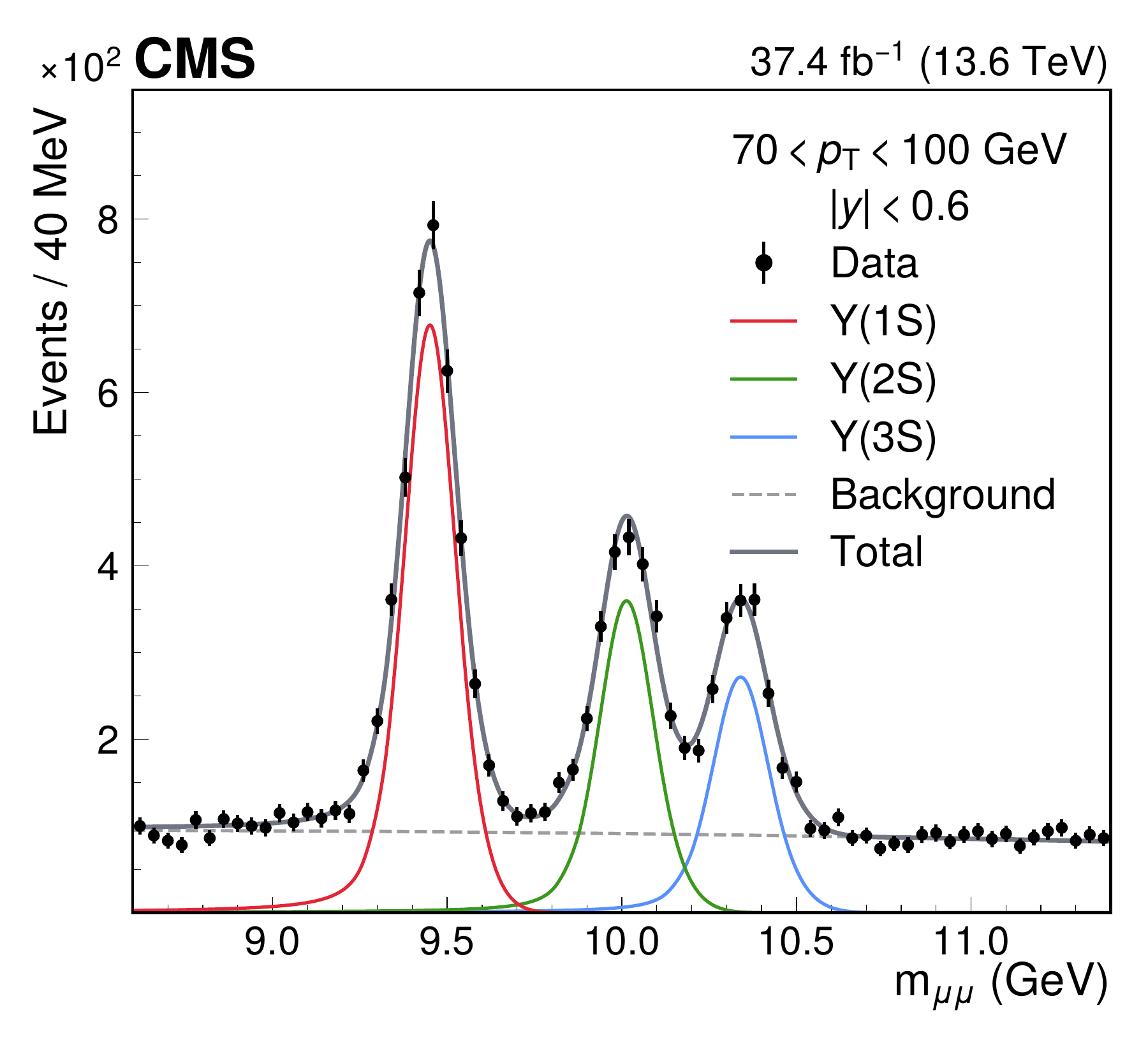}
\includegraphics[width=0.46\linewidth]{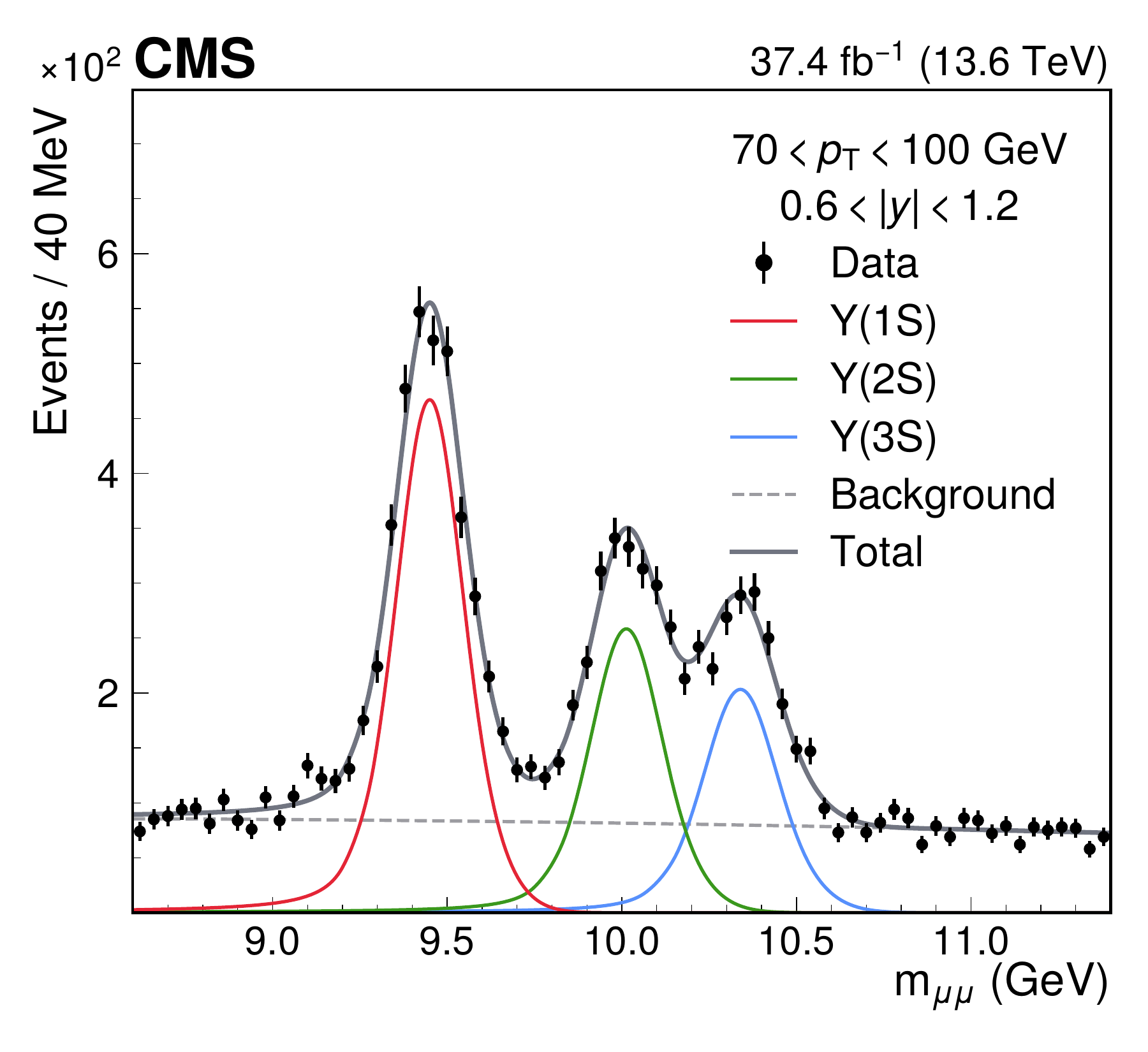}
\caption{Invariant dimuon mass distributions for dimuon 
\pt of $22 < \pt < 24 $\GeV (upper) and $70 < \pt < 100 $\GeV (lower) 
with dimuon rapidity of $\abs{y} < 0.6$ (left) and $0.6 < \abs{y} < 1.2$ (right).
The results of the fits are represented by the solid coloured lines (signal peaks), 
the dashed gray lines (background), and the solid black lines (total).}
\label{fig:massfit}
\end{figure}

A good description of the mass spectra is obtained with a relatively small number of free parameters.
Specifically, the fit model imposes
$\mu_2 = \mu_1$, $\sigma_2 / \sigma_1 = 1.55$, $n_1 = 1$, $n_2 = 2$, $\alpha_1 = \alpha_2 = 2$.
The two CB functions contribute to the total signal function as 
$f \, CB_1 + (1-f) \, CB_2$, 
where $f=0.4$ is the weight of the narrowest term.
The shapes of the three signal peaks are jointly determined,
imposing that the differences between the three $\mu$ parameters  
are identical to the differences between the world-average values of the 
\UpsilonOneS, \UpsilonTwoS, and \UpsilonThreeS masses~\cite{PDG}
and that the three $\sigma_1$ values scale in the same proportions as those masses.
In short, the free parameters in the fits, for each $(\pt,\abs{y})$ bin, 
are the amplitudes of the three signal terms,
the $\mu_1$ and $\sigma_1$ shape parameters of the \UpsilonOneS peak,
and the three parameters specifying the background term.
All the independently-fitted dimuon mass distributions are well described by this model, 
with the $\mu_1$ parameter being close to 9.45\GeV, independently of \pt and $y$, 
while $\sigma_1$ increases approximately linearly with \pt,
in the 20--200\GeV range,
from 50 to 65\MeV (from 65 to 90\MeV) for $\abs{y} < 0.6$ ($0.6 < \abs{y} < 1.2$).
Figure~\ref{fig:massfit} shows the $m_{\PGm\PGm}$ distributions measured
for two illustrative \pt bins, in both $\abs{y}$ bins,
together with the result of the corresponding fits.

\section{Systematic uncertainties}

Several sources of systematic uncertainty are considered. 
The integrated luminosity is measured with an uncertainty of 1.4\%~\cite{CMS-LUM-22-001}.
This is a global uncertainty, commonly affecting all $(\pt, y)$ bins.

The uncertainties in the signal yields are evaluated by refitting the 
$m_{\PGm\PGm}$ distributions with alternative fit models for the signal and background terms 
that still provide a reasonably good description the data.
Specifically, the fits were redone in the following alternate conditions: 
leaving the three $\mu$ parameters free;
leaving the three $\sigma_1$ parameters free;
changing the $n_1$ and $n_2$ fixed CB tail parameters by $\pm 0.5$, 
independently of each other or simultaneously;
and varying the fixed $\sigma_2 / \sigma_1$ ratio by 0.05.
To evaluate the systematic uncertainty in the signal yields 
reflecting the function assumed for the combinatorial background term, 
the fits were redone replacing the second-order polynomial 
by an exponential function, a linear function or a third-order polynomial.
For all variations, the maximal difference between the baseline and the alternate options 
(in each $(\pt,\abs{y})$ bin) is taken as the uncertainty.
The systematic uncertainty in the measured cross sections resulting from the evaluation of the signal yields
is uncorrelated in \pt and $\abs{y}$, with values in the ranges 
$\approx$\,2--3\% for the \UpsilonOneS,
$\approx$\,3--5\% for the \UpsilonTwoS, and
$\approx$\,3--7\% for the \UpsilonThreeS.

The uncertainty in the acceptance
is evaluated from the differences between the baseline values
and those obtained with a simulation where the \UpsilonOneS \pt and $\abs{y}$ distributions
are generated without any reweighting to match the simulated and measured spectra.
The corresponding uncertainty in the measured cross sections
decreases as the dimuon \pt increases,
from $\approx$\,10\% at low \pt to less than 1\% for $\pt > 50$\GeV.
This systematic uncertainty is strongly correlated between \pt and $\abs{y}$ bins.
We also verified that the differences between
the three \UpsilonnS acceptances are negligible.

The uncertainty in the dimuon detection efficiency
includes two sources:
the single muon efficiencies of each of the two muons 
and the muon-pair detection inefficiency 
occurring when the two muons have trajectories that are too close to each other.
The uncertainties in the single muon efficiencies are evaluated by 
propagating the uncertainties of the tag-and-probe measurements.
The corresponding uncertainties in the measured cross sections
vary between $\approx$\,4\% at low \pt and $\approx$\,6\% for the highest \pt bins.
The uncertainty in the factor that corrects for the muon-pair trigger inefficiency
reflects potential residual differences between the effect induced by the detector and 
its emulation in the detailed MC simulation.
To evaluate the magnitude of such potential variations, 
we start by comparing the MC event distributions (at the level of reconstructed variables)
before and after applying the dimuon trigger criteria, 
to quantify how the dimuon trigger efficiency decreases as the two muon trajectories
approach each other~\cite{BPH-10-004, BPH-13-003, BPH-14-001, BPH-22-009}.
Such studies allow us to identify events with low dimuon trigger efficiency,
those for which the inefficiency correction is the largest and, hence,
make the analysis the most sensitive to the accuracy of the MC simulation.
The cross section measurement is then redone after rejecting that subsample of events,
removing this source of a potential bias at the cost of increasing the statistical uncertainty.
The differences between the baseline cross sections and the values 
obtained after rejecting events with low dimuon detection efficiency
are within the fluctuations expected from the reduction in the size of the event sample.
Nevertheless, to ensure that the results account for 
potential inaccuracies in the MC dimuon trigger emulation,
we evaluated a corresponding uncertainty,
increasing from $\approx$\,1\% (2.5\%) in the lowest \pt bin 
to $\approx$\,6\% (9\%) for $\pt \approx 50$\GeV,
in the $\abs{y} < 0.6$ ($0.6 < \abs{y} < 1.2$) range, 
remaining flat at higher \pt.
This systematic uncertainty is strongly correlated between \pt and $\abs{y}$ bins.

\section{Results}

\begin{figure}[t!]
\centering
\includegraphics[width=0.55\linewidth]{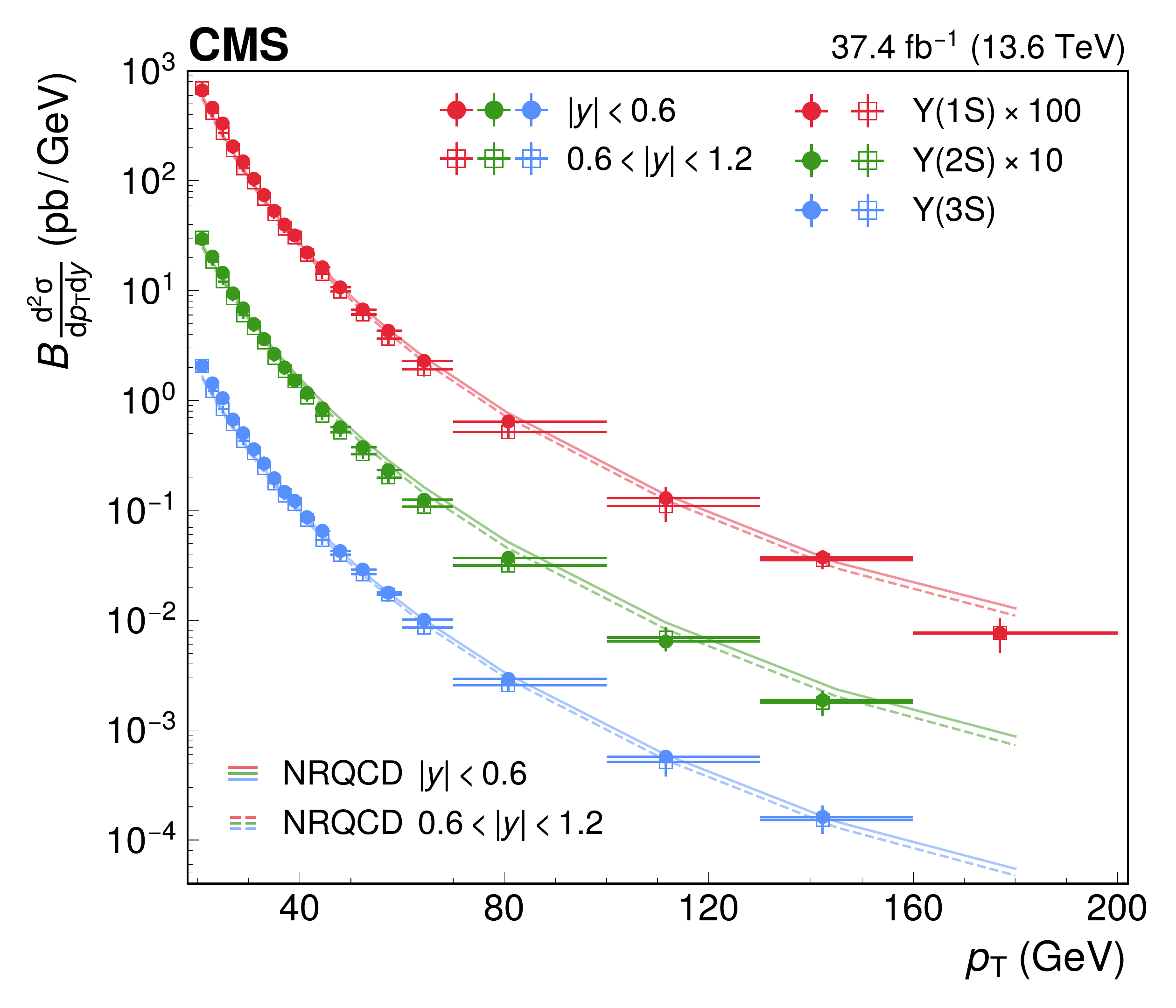}
\includegraphics[width=0.55\linewidth]{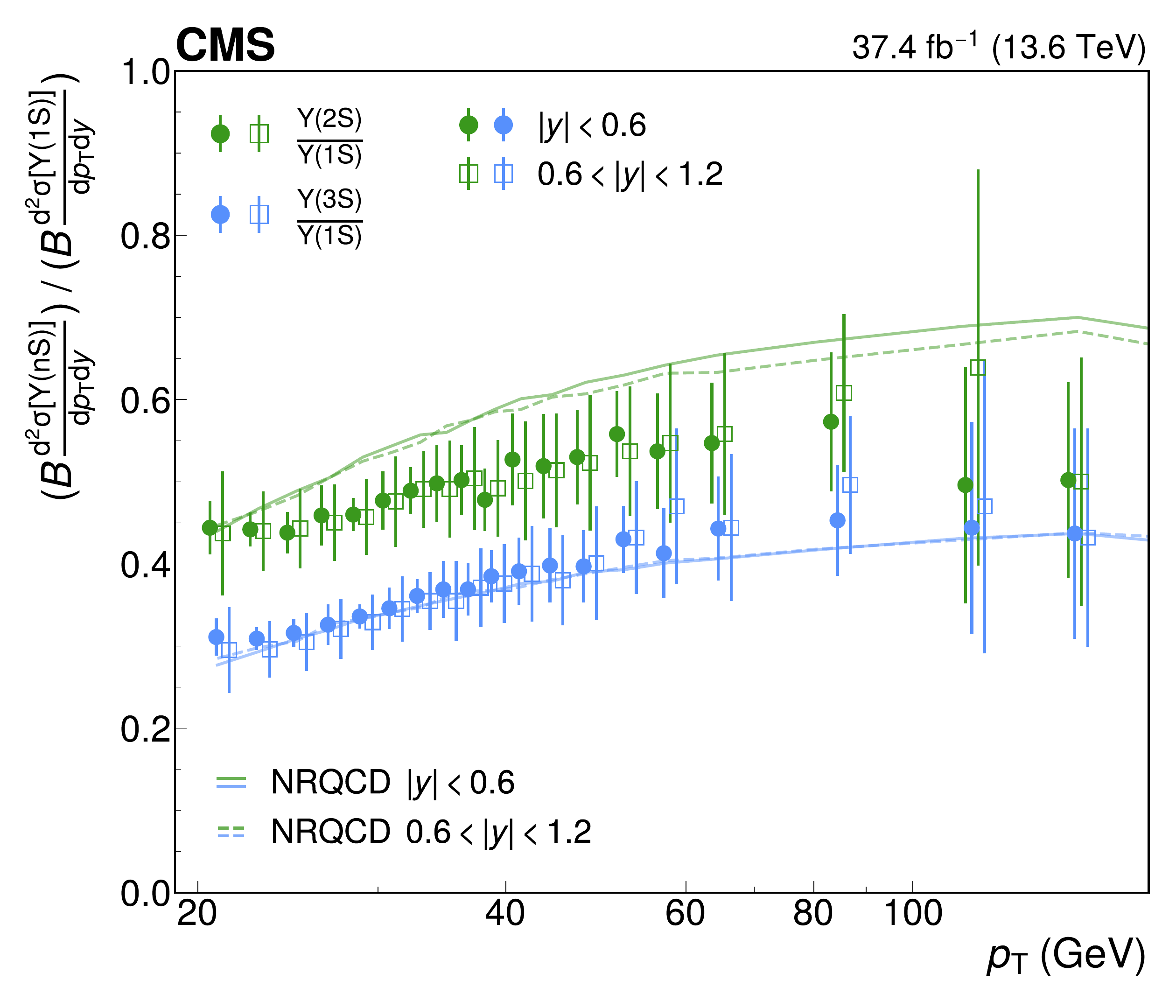}
\caption{Top: Measured \pt-differential 
\UpsilonOneS (red), \UpsilonTwoS (green), and \UpsilonThreeS (blue) cross sections
times the $\PGmp\PGmm$ decay branching fractions, per rapidity unit, 
in the 0--0.6 (filled circles) and 0.6--1.2 (open squares) $\abs{y}$ ranges. 
For visibility, the \UpsilonOneS and \UpsilonTwoS values are scaled by factors of 100 and 10, respectively.
Bottom: Corresponding \UpsilonTwoS over \UpsilonOneS (green) and \UpsilonThreeS over \UpsilonOneS (blue) ratios,
in the 0--0.6 (filled circles) and 0.6--1.2 (open squares) $\abs{y}$ ranges.
For visibility purposes, the data points are slightly displaced in the $x$ axis.
In both panels, the error bars represent the total uncertainties.
The NRQCD curves are shown for illustration.}
\label{fig:cross_sections-and-ratios}
\end{figure}

Figure~\ref{fig:cross_sections-and-ratios}~(top) shows
the measured \pt-differential \UpsilonnS cross sections
times the branching fractions of their $\PGmp\PGmm$ decays
for $\abs{y} < 0.6$ and $0.6 < \abs{y} < 1.2$.
The data points are shown at the mean values of the \pt distributions of the measured events inside each \pt bin;
the numerical values, for the three \UpsilonnS states,
are listed in Tables~\ref{tab:xs-1S}--\ref{tab:xs-3S}, in Appendix~\ref{app:tables}.
The measured cross sections include feed-down decays from heavier bottomonium states 
and are based on the assumption that the mesons are produced unpolarized. 
Table~\ref{tab:polarization-factors} can be used to adjust the data to other polarization scenarios.
The two rapidity ranges (both close to midrapidity) lead to almost identical cross sections.
No \UpsilonTwoS and \UpsilonThreeS values are provided for the highest \pt bin 
because the collected number of events is too small to provide a robust extraction of those two yields.

In order to better appreciate how the shapes of the \pt-differential cross sections
vary among the three \UpsilonnS states, 
Fig.~\ref{fig:cross_sections-and-ratios}~(bottom) shows the cross section ratios with respect to the ground state.
For $\pt \gtrsim 55$\GeV, the ratios seem to reach a plateau.

Both panels of Fig.~\ref{fig:cross_sections-and-ratios} 
include NRQCD predictions~\cite{Feng:2020cvm}
for the collision energy and kinematical phase space of our measurement.
Contributions from feed-down decays of heavier S- and P-wave bottomonia are included. 
The NRQCD predictions describe the general trends of the measured cross sections reasonably well. 
Different curves would be obtained if computed with alternative LDME values, 
with other feed-down fractions and polarization scenarios, 
including higher-order terms in the perturbative expansion, etc., so that 
this specific comparison 
is only meant to illustrate that the new measurements reported in this article 
provide important constraints for improved phenomenological global-fit data analyses.

\begin{figure}[h!]
\centering  
\includegraphics[width=0.5\linewidth]{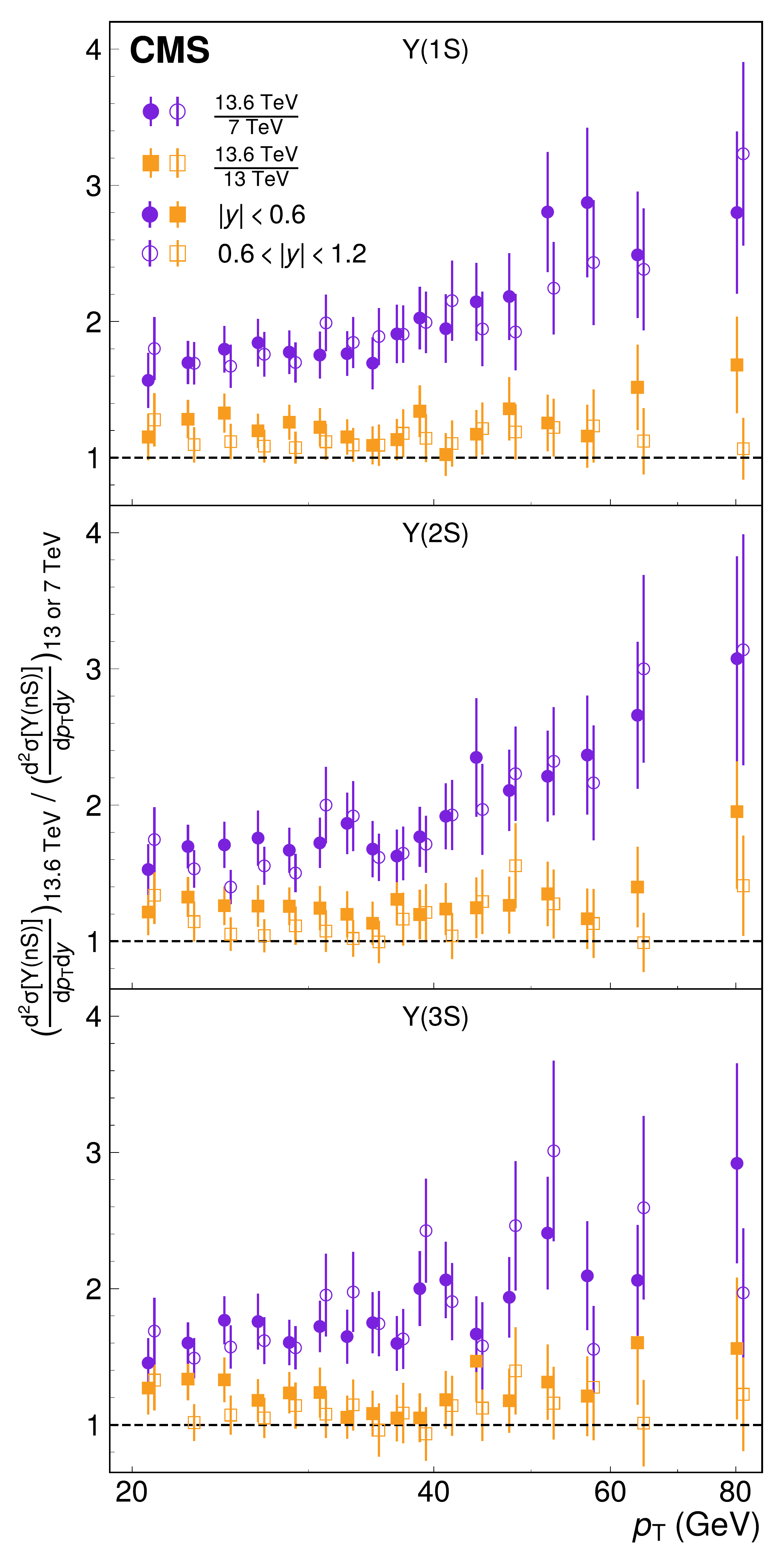}
\caption{Ratios of \UpsilonnS cross sections between this measurement, made at $\sqrt{s} = 13.6$\TeV, 
and previous measurements, made at 7 and 13\TeV~\cite{CMS:2015xqv,CMS13TeV},
treating all uncertainties as uncorrelated.
For visibility purposes, the data points are slightly displaced in the $x$ axis.}
\label{fig:energy-ratios}
\end{figure}

Figure~\ref{fig:energy-ratios} shows
the ratios between the new \UpsilonnS measurements, made at $\sqrt{s}=13.6$\TeV,
and the corresponding cross sections measured at 7 and 13\TeV,
using event samples corresponding to integrated luminosities of 4.9 and 2.7\fbinv, 
respectively~\cite{CMS:2015xqv,CMS13TeV}. 

\section{Summary}

The cross sections of the \UpsilonOneS, \UpsilonTwoS, and \UpsilonThreeS bottomonium states,
times the branching fractions of their dimuon decays, were measured 
as a function of their transverse momentum in a broad \pt range, 
from 20 to 200\GeV, in two rapidity regions, $\abs{y} < 0.6$ and $0.6 < \abs{y} < 1.2$.
The analysis was based on an event sample of $\Pp\Pp$ collisions 
at $\sqrt{s} =13.6$\TeV, collected by CMS in 2022 and corresponding to an integrated luminosity of 37.4\fbinv.
The reported cross sections include feed-down contributions from heavier bottomonia.
These results are an important input to global fits of quarkonium cross sections and polarizations,
made in the context of the NRQCD framework or of other models of quarkonium production.

\begin{acknowledgments}
We congratulate our colleagues in the CERN accelerator departments for the excellent performance of the LHC and thank the technical and administrative staffs at CERN and at other CMS institutes for their contributions to the success of the CMS effort. In addition, we gratefully acknowledge the computing centres and personnel of the Worldwide LHC Computing Grid and other centres for delivering so effectively the computing infrastructure essential to our analyses. Finally, we acknowledge the enduring support for the construction and operation of the LHC, the CMS detector, and the supporting computing infrastructure provided by the following funding agencies: SC (Armenia), BMBWF and FWF (Austria); FNRS and FWO (Belgium); CNPq, CAPES, FAPERJ, FAPERGS, and FAPESP (Brazil); MES and BNSF (Bulgaria); CERN; CAS, MoST, and NSFC (China); MINCIENCIAS (Colombia); MSES and CSF (Croatia); RIF (Cyprus); SENESCYT (Ecuador); ERC PRG, TARISTU24-TK10 and MoER TK202 (Estonia); Academy of Finland, MEC, and HIP (Finland); CEA and CNRS/IN2P3 (France); SRNSF (Georgia); BMFTR, DFG, and HGF (Germany); GSRI (Greece); NKFIH (Hungary); DAE and DST (India); IPM (Iran); SFI (Ireland); INFN (Italy); MSIT and NRF (Republic of Korea); MES (Latvia); LMTLT (Lithuania); MOE and UM (Malaysia); BUAP, CINVESTAV, CONACYT, LNS, SEP, and UASLP-FAI (Mexico); MOS (Montenegro); MBIE (New Zealand); PAEC (Pakistan); MES, NSC, and NAWA (Poland); FCT (Portugal); MESTD (Serbia); MICIU/AEI and PCTI (Spain); MOSTR (Sri Lanka); Swiss Funding Agencies (Switzerland); MST (Taipei); MHESI (Thailand); TUBITAK and TENMAK (T\"{u}rkiye); NASU (Ukraine); STFC (United Kingdom); DOE and NSF (USA).

\hyphenation{Rachada-pisek} Individuals have received support from the Marie-Curie programme and the European Research Council and Horizon 2020 Grant, contract Nos.\ 675440, 724704, 752730, 758316, 765710, 824093, 101115353, 101002207, 101001205, and COST Action CA16108 (European Union); the Leventis Foundation; the Alfred P.\ Sloan Foundation; the Alexander von Humboldt Foundation; the Science Committee, project no. 22rl-037 (Armenia); the Fonds pour la Formation \`a la Recherche dans l'Industrie et dans l'Agriculture (FRIA) and Fonds voor Wetenschappelijk Onderzoek contract No. 1228724N (Belgium); the Beijing Municipal Science \& Technology Commission, No. Z191100007219010, the Fundamental Research Funds for the Central Universities, the Ministry of Science and Technology of China under Grant No. 2023YFA1605804, the Natural Science Foundation of China under Grant No. 12535004, and USTC Research Funds of the Double First-Class Initiative No.\ YD2030002017 (China); the Ministry of Education, Youth and Sports (MEYS) of the Czech Republic; the Shota Rustaveli National Science Foundation, grant FR-22-985 (Georgia); the Deutsche Forschungsgemeinschaft (DFG), among others, under Germany's Excellence Strategy -- EXC 2121 ``Quantum Universe" -- 390833306, and under project number 400140256 - GRK2497; the Hellenic Foundation for Research and Innovation (HFRI), Project Number 2288 (Greece); the Hungarian Academy of Sciences, the New National Excellence Program - \'UNKP, the NKFIH research grants K 131991, K 133046, K 138136, K 143460, K 143477, K 146913, K 146914, K 147048, 2020-2.2.1-ED-2021-00181, TKP2021-NKTA-64, and 2025-1.1.5-NEMZ\_KI-2025-00004 (Hungary); the Council of Science and Industrial Research, India; ICSC -- National Research Centre for High Performance Computing, Big Data and Quantum Computing, FAIR -- Future Artificial Intelligence Research, and CUP I53D23001070006 (Mission 4 Component 1), funded by the NextGenerationEU program (Italy); the Latvian Council of Science; the Ministry of Education and Science, project no. 2022/WK/14, and the National Science Center, contracts Opus 2021/41/B/ST2/01369, 2021/43/B/ST2/01552, 2023/49/B/ST2/03273, and the NAWA contract BPN/PPO/2021/1/00011 (Poland); the Funda\c{c}\~ao para a Ci\^encia e a Tecnologia, grant CEECIND/01334/2018 (Portugal); the National Priorities Research Program by Qatar National Research Fund; MICIU/AEI/10.13039/501100011033, ERDF/EU, ``European Union NextGenerationEU/PRTR", and Programa Severo Ochoa del Principado de Asturias (Spain); the Chulalongkorn Academic into Its 2nd Century Project Advancement Project, the National Science, Research and Innovation Fund program IND\_FF\_68\_369\_2300\_097, and the Program Management Unit for Human Resources \& Institutional Development, Research and Innovation, grant B39G680009 (Thailand); the Kavli Foundation; the Nvidia Corporation; the SuperMicro Corporation; the Welch Foundation, contract C-1845; and the Weston Havens Foundation (USA).    

\end{acknowledgments}

\bibliography{auto_generated}

\appendix
\numberwithin{table}{section}
\section{Tables of cross sections and correction factors}
\label{app:tables}

This appendix collects 
a table (\ref{tab:polarization-factors}) listing the scaling factors needed to derive,
from the ``unpolarized values",
the cross sections corresponding to any other polarization scenario,
by interpolating between the two most extreme polarization cases, in the helicity frame. 
It also includes three tables (\ref{tab:xs-1S}--\ref{tab:xs-3S})
with the numerical values of the \pt-differential
\UpsilonOneS, \UpsilonTwoS, and \UpsilonThreeS cross section measurements,
in two rapidity ranges, under the assumption of unpolarized production. 
The $\langle \pt \rangle$ values are the mean values of the 
background-subtracted and efficiency-corrected \pt distributions, in each \pt bin.

\begin{table}[htbp]
\centering
\topcaption{Scaling factors needed to obtain the $\Upsilon$($n$S) differential cross sections 
corresponding to two extreme polarization scenarios in the helicity frame, 
$\lambda_{\theta}^{\mathrm{HX}} = +1$ or $-1$, from the values reported in Tables~\ref{tab:xs-1S}--\ref{tab:xs-3S}.}
\label{tab:polarization-factors}
\begin{tabular}{*{5}{c}}
\rule{0pt}{0.4cm}$\Delta \pt$ & \multicolumn{2}{c}{$\abs{y} < 0.6$} & \multicolumn{2}{c}{$0.6 < \abs{y} < 1.2$} \\
\cline{2-5}
\rule{0pt}{2.2ex}\hfil (\GeVns{}) \hfil & 
$\lambda_{\theta} = +1$ & $\lambda_{\theta} = -1$ & $\lambda_{\theta} = +1$ & $\lambda_{\theta} = -1$ \\ 
\hline
20--22	&	1.32	&	0.67	&	1.30	&	0.68	\\
22--24	&	1.31	&	0.68	&	1.28	&	0.70	\\
24--26	&	1.30	&	0.68	&	1.27	&	0.70	\\
26--28	&	1.29	&	0.69	&	1.27	&	0.70	\\
28--30	&	1.28	&	0.69	&	1.25	&	0.71	\\
30--32	&	1.26	&	0.70	&	1.25	&	0.71	\\
32--34	&	1.26	&	0.71	&	1.25	&	0.72	\\
34--36	&	1.25	&	0.72	&	1.23	&	0.73	\\
36--38	&	1.24	&	0.72	&	1.23	&	0.73	\\
38--40	&	1.23	&	0.73	&	1.22	&	0.74	\\
40--43	&	1.22	&	0.73	&	1.21	&	0.74	\\
43--46	&	1.21	&	0.75	&	1.20	&	0.75	\\
46--50	&	1.20	&	0.75	&	1.19	&	0.76	\\
50--55	&	1.18	&	0.77	&	1.18	&	0.77	\\
55--60	&	1.17	&	0.78	&	1.16	&	0.78	\\
60--70	&	1.15	&	0.79	&	1.15	&	0.80	\\
70--100	&	1.12	&	0.83	&	1.12	&	0.83	\\
100--130	&	1.09	&	0.86	&	1.09	&	0.86	\\
130--160	&	1.07	&	0.88	&	1.07	&	0.88	\\
160--200	&	1.06	&	0.90	&		&		\\
\end{tabular}
\end{table}

\begin{table}[htb]
\centering
\topcaption{The $\Upsilon$(1S) differential cross section times dimuon branching fraction,
$\mathcal{B}\, \rd^2\sigma/\rd{}\pt\rd{}y$, in two rapidity ranges, for the unpolarized scenario.
The relative uncertainties (statistical and systematic) are given in percent.
The systematic uncertainties are partially correlated across all the $(\pt,\abs{y})$ bins.}
\label{tab:xs-1S}
\begin{tabular}{cc ccc ccc}
\rule{0pt}{0.4cm} & & \multicolumn{6}{c}{$\mathcal{B}\, \rd^2\sigma/\rd{}\pt\rd{}y$} \\ 
\cline{3-8}
\rule{0pt}{0.4cm} $\Delta \pt$ & $\langle \pt \rangle$ & \multicolumn{3}{c}{$\abs{y} < 0.6$} & \multicolumn{3}{c}{$0.6 < \abs{y} < 1.2$} \\
\cline{3-8}
\rule{0pt}{0.4cm} \hfil (\GeVns{}) \hfil & \hfil (\GeVns{}) \hfil 
& (pb/\GeVns{}) & $\delta_{\text{stat}}$ (\%) & $\delta_{\text{syst}}$ (\%) 
& (pb/\GeVns{}) & $\delta_{\text{stat}}$ (\%) & $\delta_{\text{syst}}$ (\%) \\
\hline
20--22	&	20.9 	& $6.64$	&	0.50 &	11	&	$6.98$ &	0.58 &	12	\\
22--24	&	22.9 	& $4.62$	&	0.47 &	7.4	&	$4.13$ &	0.54 &	8.0	\\
24--26	&	24.9 	& $3.32$	&	0.50 &	7.6	&	$2.73$ &	0.56 &	8.2	\\
26--28	&	26.9 	& $2.06$	&	0.54 &	8.5	&	$1.90$ &	0.61 &	7.9	\\
28--30	&	28.9 	& $1.50$	&	0.61 &	6.7	&	$1.30$ &	0.67 &	7.1	\\
30--32	&	31.0 	& $1.04$	&	0.69 &	7.8	&	$9.63 \times 10^{-1}$ &	0.75 &	8.0	\\
32--34	&	33.0 	& $7.41 \times 10^{-1}$	&	0.77 &	7.3	&	$6.85 \times 10^{-1}$ &	0.84 &	6.9	\\
34--36	&	35.0 	& $5.32 \times 10^{-1}$	&	0.88 &	8.5	&	$4.97 \times 10^{-1}$ &	0.95 &	9.3	\\
36--38	&	37.0 	& $3.99 \times 10^{-1}$	&	0.99 &	8.6	&	$3.68 \times 10^{-1}$ &	1.1 &	8.9	\\
38--40	&	39.0 	& $3.18 \times 10^{-1}$	&	1.1 &	8.4	&	$3.03 \times 10^{-1}$ &	1.2 &	8.8	\\
40--43	&	41.4 	& $2.22 \times 10^{-1}$	&	1.1 &	8.3	&	$2.11 \times 10^{-1}$ &	1.1 &	11	\\
43--46	&	44.4 	& $1.63 \times 10^{-1}$	&	1.2 &	11	&	$1.42 \times 10^{-1}$ &	1.3 &	10	\\
46--50	&	47.9 	& $1.07 \times 10^{-1}$	&	1.3 &	8.9	&	$9.81 \times 10^{-2}$ &	1.4 &	12	\\
50--55	&	52.3 	& $6.73 \times 10^{-2}$	&	1.5 &	9.1	&	$6.06 \times 10^{-2}$ &	1.6 &	11	\\
55--60	&	57.3 	& $4.31 \times 10^{-2}$	&	1.8 &	9.5	&	$3.65 \times 10^{-2}$ &	2.0 &	14	\\
60--70	&	64.3 	& $2.29 \times 10^{-2}$	&	1.8 &	11	&	$1.93 \times 10^{-2}$ &	2.0 &	14	\\
70--100	&	80.8 	& $6.44 \times 10^{-3}$	&	2.1 &	14	&	$5.17 \times 10^{-3}$ &	2.3 &	12	\\
100--130	&	111.6 	& $1.29 \times 10^{-3}$	&	5.1 &	12	&	$1.09 \times 10^{-3}$ &	5.8 &	27	\\
130--160	&	142.3 	& $3.72 \times 10^{-4}$	&	11 &	13	&	$3.53 \times 10^{-4}$ &	11 &	13	\\
160--200	&	177.0 	& $7.61 \times 10^{-5}$	&	26 &	9.8	&	$7.71 \times 10^{-5}$ &	29 &	19	\\
\end{tabular}
\end{table}

\begin{table}[htb]
\centering
\topcaption{The $\Upsilon$(2S) differential cross section times dimuon branching fraction,
$\mathcal{B}\, \rd^2\sigma/\rd{}\pt\rd{}y$, in two rapidity ranges, for the unpolarized scenario.
The relative uncertainties (statistical and systematic) are given in percent.
The systematic uncertainties are partially correlated across all the $(\pt,\abs{y})$ bins.}
\label{tab:xs-2S}
\begin{tabular}{cc ccc ccc}
\rule{0pt}{0.4cm} & & \multicolumn{6}{c}{$\mathcal{B}\, \rd^2\sigma/\rd{}\pt\rd{}y$} \\ 
\cline{3-8}
\rule{0pt}{0.4cm} $\Delta \pt$ & $\langle \pt \rangle$ & \multicolumn{3}{c}{$\abs{y} < 0.6$} & \multicolumn{3}{c}{$0.6 < \abs{y} < 1.2$} \\
\cline{3-8}
\rule{0pt}{0.4cm} \hfil 
(\GeVns{}) \hfil & \hfil (\GeVns{}) \hfil 
& (pb/\GeVns{}) & $\delta_{\text{stat}}$ (\%) & $\delta_{\text{syst}}$ (\%) 
& (pb/\GeVns{}) & $\delta_{\text{stat}}$ (\%) & $\delta_{\text{syst}}$ (\%) \\
\hline
20--22	&	20.9	& $2.95$	&	0.80 &	11	&	$3.05$ &	0.95 &	13	\\
22--24	&	22.9	& $2.04$	&	0.75 &	7.6	&	$1.82$ &	0.88 &	8.5	\\
24--26	&	24.9	& $1.45$	&	0.79 &	8.0	&	$1.21$ &	0.91 &	8.2	\\
26--28	&	26.9	& $9.44 \times 10^{-1}$	&	0.84 &	9.0	&	$8.56 \times 10^{-1}$ &	1.0 &	8.6	\\
28--30	&	28.9	& $6.89 \times 10^{-1}$	&	0.94 &	6.7	&	$5.96 \times 10^{-1}$ &	1.1 &	7.5	\\
30--32	&	31.0	& $4.96 \times 10^{-1}$	&	1.0 &	8.1	&	$4.58 \times 10^{-1}$ &	1.2 &	8.4	\\
32--34	&	33.0	& $3.62 \times 10^{-1}$	&	1.2 &	7.7	&	$3.36 \times 10^{-1}$ &	1.3 &	6.9	\\
34--36	&	35.0	& $2.65 \times 10^{-1}$	&	1.3 &	8.8	&	$2.44 \times 10^{-1}$ &	1.5 &	10	\\
36--38	&	37.0	& $2.00 \times 10^{-1}$	&	1.5 &	8.4	&	$1.86 \times 10^{-1}$ &	1.6 &	9.3	\\
38--40	&	39.0	& $1.52 \times 10^{-1}$	&	1.7 &	8.2	&	$1.49 \times 10^{-1}$ &	1.9 &	9.1	\\
40--43	&	41.4	& $1.17 \times 10^{-1}$	&	1.5 &	8.7	&	$1.06 \times 10^{-1}$ &	1.7 &	10	\\
43--46	&	44.4	& $8.46 \times 10^{-2}$	&	1.8 &	12	&	$7.28 \times 10^{-2}$ &	2.0 &	10	\\
46--50	&	47.9	& $5.69 \times 10^{-2}$	&	1.9 &	8.8	&	$5.13 \times 10^{-2}$ &	2.1 &	13	\\
50--55	&	52.3	& $3.76 \times 10^{-2}$	&	2.1 &	8.8	&	$3.25 \times 10^{-2}$ &	2.3 &	11	\\
55--60	&	57.3	& $2.32 \times 10^{-2}$	&	2.7 &	9.3	&	$1.99 \times 10^{-2}$ &	3.0 &	15	\\
60--70	&	64.3	& $1.25 \times 10^{-2}$	&	2.6 &	10	&	$1.08 \times 10^{-2}$ &	2.9 &	14	\\
70--100	&	80.5	& $3.69 \times 10^{-3}$	&	3.0 &	14	&	$3.14 \times 10^{-3}$ &	3.3 &	11	\\
100--130	&	111.6	& $6.39 \times 10^{-4}$	&	8.3 &	13	&	$6.97 \times 10^{-4}$ &	8.2 &	21	\\
130--160	&	142.3	& $1.87 \times 10^{-4}$	&	19 &	9.4	&	$1.77 \times 10^{-4}$ &	19 &	14	\\
\end{tabular}
\end{table}

\begin{table}[htb]
\centering
\topcaption{The $\Upsilon$(3S) differential cross section times dimuon branching fraction,
$\mathcal{B}\, \rd^2\sigma/\rd{}\pt\rd{}y$, in two rapidity ranges, for the unpolarized scenario.
The relative uncertainties (statistical and systematic) are given in percent.
The systematic uncertainties are partially correlated across all the $(\pt,\abs{y})$ bins.}
\label{tab:xs-3S}
\begin{tabular}{cc ccc ccc}
\rule{0pt}{0.4cm} & & \multicolumn{6}{c}{$\mathcal{B}\, \rd^2\sigma/\rd{}\pt\rd{}y$} \\ 
\cline{3-8}
\rule{0pt}{0.4cm} $\Delta \pt$ & $\langle \pt \rangle$ & \multicolumn{3}{c}{$\abs{y} < 0.6$} & \multicolumn{3}{c}{$0.6 < \abs{y} < 1.2$} \\
\cline{3-8}
\rule{0pt}{0.4cm} \hfil 
(\GeVns{}) \hfil & \hfil (\GeVns{}) \hfil 
& (pb/\GeVns{}) & $\delta_{\text{stat}}$ (\%) & $\delta_{\text{syst}}$ (\%) 
& (pb/\GeVns{}) & $\delta_{\text{stat}}$ (\%) & $\delta_{\text{syst}}$ (\%) \\
\hline
20--22	&	20.9	& $2.07$	&	0.50 &	11	&	$2.06$ &	1.2 &	13	\\
22--24	&	22.9	& $1.43$	&	0.58 &	7.5	&	$1.22$ &	1.1 &	8.2	\\
24--26	&	24.9	& $1.05$	&	0.69 &	7.8	&	$8.33 \times 10^{-1}$ &	1.1 &	8.2	\\
26--28	&	26.9	& $6.70 \times 10^{-1}$	&	0.86 &	8.7	&	$6.10 \times 10^{-1}$ &	1.2 &	8.2	\\
28--30	&	28.9	& $5.04 \times 10^{-1}$	&	0.94 &	6.7	&	$4.29 \times 10^{-1}$ &	1.3 &	7.3	\\
30--32	&	31.0	& $3.60 \times 10^{-1}$	&	1.1 &	8.0	&	$3.32 \times 10^{-1}$ &	1.4 &	8.2	\\
32--34	&	33.0	& $2.67 \times 10^{-1}$	&	1.3 &	7.5	&	$2.43 \times 10^{-1}$ &	1.6 &	6.9	\\
34--36	&	35.0	& $1.96 \times 10^{-1}$	&	1.4 &	8.7	&	$1.76 \times 10^{-1}$ &	1.8 &	9.8	\\
36--38	&	37.0	& $1.47 \times 10^{-1}$	&	1.8 &	8.5	&	$1.37 \times 10^{-1}$ &	2.0 &	9.1	\\
38--40	&	39.0	& $1.22 \times 10^{-1}$	&	1.8 &	8.4	&	$1.14 \times 10^{-1}$ &	2.2 &	8.9	\\
40--43	&	41.4	& $8.67 \times 10^{-2}$	&	1.7 &	8.5	&	$8.19 \times 10^{-2}$ &	2.0 &	10	\\
43--46	&	44.4	& $6.50 \times 10^{-2}$	&	2.1 &	11	&	$5.37 \times 10^{-2}$ &	2.5 &	10	\\
46--50	&	47.9	& $4.26 \times 10^{-2}$	&	2.5 &	8.9	&	$3.94 \times 10^{-2}$ &	2.5 &	12	\\
50--55	&	52.3	& $2.89 \times 10^{-2}$	&	2.4 &	8.9	&	$2.62 \times 10^{-2}$ &	2.7 &	11	\\
55--60	&	57.3	& $1.78 \times 10^{-2}$	&	3.2 &	9.4	&	$1.71 \times 10^{-2}$ &	3.3 &	14	\\
60--70	&	64.3	& $1.01 \times 10^{-2}$	&	3.0 &	11	&	$8.56 \times 10^{-3}$ &	3.4 &	14	\\
70--100	&	80.6	& $2.92 \times 10^{-3}$	&	3.8 &	14	&	$2.56 \times 10^{-3}$ &	3.8 &	11	\\
100--130	&	111.6	& $5.72 \times 10^{-4}$	&	9.7 &	12	&	$5.13 \times 10^{-4}$ &	10 &	24	\\
130--160	&	142.3	& $1.62 \times 10^{-4}$	&	25 &	11	&	$1.52 \times 10^{-4}$ &	22 &	13	\\
\end{tabular}
\end{table}

\cleardoublepage \section{The CMS Collaboration \label{app:collab}}\begin{sloppypar}\hyphenpenalty=5000\widowpenalty=500\clubpenalty=5000\cmsinstitute{Yerevan Physics Institute, Yerevan, Armenia}
{\tolerance=6000
A.~Hayrapetyan, V.~Makarenko\cmsorcid{0000-0002-8406-8605}, A.~Tumasyan\cmsAuthorMark{1}\cmsorcid{0009-0000-0684-6742}
\par}
\cmsinstitute{Institut f\"{u}r Hochenergiephysik, Vienna, Austria}
{\tolerance=6000
W.~Adam\cmsorcid{0000-0001-9099-4341}, L.~Benato\cmsorcid{0000-0001-5135-7489}, T.~Bergauer\cmsorcid{0000-0002-5786-0293}, M.~Dragicevic\cmsorcid{0000-0003-1967-6783}, P.S.~Hussain\cmsorcid{0000-0002-4825-5278}, M.~Jeitler\cmsAuthorMark{2}\cmsorcid{0000-0002-5141-9560}, N.~Krammer\cmsorcid{0000-0002-0548-0985}, A.~Li\cmsorcid{0000-0002-4547-116X}, D.~Liko\cmsorcid{0000-0002-3380-473X}, M.~Matthewman, J.~Schieck\cmsAuthorMark{2}\cmsorcid{0000-0002-1058-8093}, R.~Sch\"{o}fbeck\cmsAuthorMark{2}\cmsorcid{0000-0002-2332-8784}, M.~Shooshtari\cmsorcid{0009-0004-8882-4887}, M.~Sonawane\cmsorcid{0000-0003-0510-7010}, W.~Waltenberger\cmsorcid{0000-0002-6215-7228}, C.-E.~Wulz\cmsAuthorMark{2}\cmsorcid{0000-0001-9226-5812}
% ...
\par}
\vskip\cmsinstskip
\dag:~Deceased\\
$^{1}$Also at Yerevan State University, Yerevan, Armenia\\
$^{2}$Also at TU Wien, Vienna, Austria\\
% ...
\end{sloppypar}
\end{document}